\documentclass[conference,onecolumn,12pt]{IEEEtran}

\usepackage{graphicx}
\usepackage{amsmath,epsfig,amssymb}
\usepackage{times,amsmath,epsfig,amssymb,graphicx,amsfonts}
\title{Capacity Achieving Linear Codes with Random Binary Sparse Generating Matrices}
\author{A. Makhdoumi Kakhaki, H. Karkeh Abadi, P. Pad, H. Saeedi, F. Marvasti, K. Alishahi\\
Advanced Communications Research Institute, Sharif University of Technology,
Tehran, Iran\\
Email: \{makhdoumi, karkehabadi, pedram\_pad\}@ee.sharif.edu,\\ hsaeedi@ieee.org, marvasti@sharif.edu, alishahi@sharif.edu
}
\begin{document}
\maketitle
\newtheorem{remark}{\textbf{Remark}}
\newtheorem{theorem}{\textbf{Theorem}}
\newtheorem{definition}{\textbf{Definition}}
\newtheorem{example}{\textbf{Example}}
\newtheorem{corollary}{\textbf{Corollary}}
\newtheorem{lemma}{\textbf{Lemma}}
\newtheorem{note}{\textbf{Note}}
\newtheorem{proposition}{\textbf{Proposition}}
\newtheorem{conjecture}{\textbf{Conjecture}}

\begin{abstract}
In this paper, we prove the existence of capacity achieving linear codes with random binary sparse generating matrices. The results on the existence of capacity achieving linear codes in the literature are limited to the random binary codes with equal probability generating matrix elements and sparse parity-check matrices. Moreover, the codes with sparse generating matrices reported in the literature are not proved to be capacity achieving.

 As opposed to the existing results in the literature, which are based on optimal maximum a posteriori decoders, the proposed approach is based on a different decoder and consequently is suboptimal. We also demonstrate an interesting trade-off between the sparsity of the generating matrix and the error exponent (a constant which determines how exponentially fast the probability of error decays as block length tends to infinity). An interesting observation is that for small block sizes, less sparse generating matrices have better performances while for large blok sizes, the performance of the random generating matrices become independent of the sparsity. Moreover, we prove the existence of capacity achieving linear codes with a given (arbitrarily low) density of ones on rows of the generating matrix. In addition to proving the existence of capacity achieving sparse codes, an important conclusion of our paper is that for a sufficiently large code length, no search is necessary in practice to find a deterministic matrix by proving that any arbitrarily selected sequence of sparse generating matrices is capacity achieving with high probability. The focus in this paper is on the binary symmetric and binary erasure channels.
\end{abstract}


\section{Introduction}\label{sec:intro}
The Shannon coding theorem \cite{Shannon} states that for a variety of channels with a given capacity $C$, if the information transmission rate $R$ over the channel is below $C$, there exists a coding scheme for which the information can be transmitted with an arbitrarily low probability of error. For Discrete Memoryless Channels (DMC), it has been shown \cite{Fano} that the probability of error can be bounded between two exponentially decaying functions of the codeword block length, $n$. In this theorem, there is no constraint on the codes in terms of linearity. In \cite{Gal}, a simpler proof of the Shannon theorem has been provided. The existence of capacity achieving linear codes over the Binary Symmetric Channel (BSC) was shown by Elias \cite{Galbook} where it was also proved that linear codes have the same error exponent as random codes. A similar result has been obtained in \cite{Largedev}. It was recently shown in \cite{Forney} that the error exponent of a typical random linear code can, in fact, be larger than a typical random code, implying a faster decaying of error as $n$ increases.
Some bounds on the decoding error probability of linear codes have been derived in \cite{94}. The result reported in \cite{Galbook}-\cite{94} are all based on the fact that the elements of generating matrices of the capacity achieving linear codes should be one or zero with equal probability; therefore the generating matrix of such approaches are not sparse.\footnote{A sparse generating matrix is a matrix with a statistically low density of ones, see Section II for the exact definition.} Moreover, most papers on capacity achieving sparse linear codes are concentrated on codes with sparse parity-check matrices. In particular, an important class of codes called Low-Density Parity-Check (LDPC) codes \cite{ldpc,makay} have been of major interest in the past decade. While these codes have sparse parity-check matrices, they do not necessarily exhibit sparse generating matrices which are the focus of this paper.
In \cite{LDGM1}-\cite{LDGM2}, some Low-Density Generating-Matrix (LDGM) schemes have been proposed which have performance approaching the capacity.\footnote{We distinguish between ``capacity approaching'' and ``capacity achieving'' codes. The former term is used when the performance of
the code can be shown numerically to approach capacity without
any guarantee to achieve it. The latter term is used if the performance can be rigorously proved to achieve the capacity. The subject of this paper is on the latter case.} Some other related literature on the codes with sparse generating matrices having performance close to capacity includes \cite{LDGM3}-\cite{Raptor};
in \cite{LDGM3}, a capacity-achieving scheme has been proposed based on serially concatenated codes with an outer LDPC code and an inner LDGM code. However, the generating matrix corresponding to the concatenation is not necessarily sparse. On the other hand, rateless codes have been proposed in \cite{LT} and \cite{Raptor} which have sparse generating matrices but are only proved to be capacity achieving over the Binary Erasure Channel (BEC).

In this paper, using a novel approach, we prove the existence of capacity achieving linear codes with \emph{sparse generating matrices} that can provide reliable communications over two important classes of DMC channels; namely, BEC and BSC at rates below the channel capacity. The proof is accomplished by first deriving a lower bound on the probability of correct detection for a given generating matrix and then by taking the expectation of that lower bound over all possible generating matrices with elements 1 and 0 with probability $\rho$ and $1-\rho$, respectively. By showing that this expectation goes to one as $n$ approaches infinity, we prove the existence of linear capacity achieving codes. To show the sparsity, we extend this result by taking the expectation over a subset of matrices for which the density of ones could be made arbitrarily close to any target $\rho$. We then prove a stronger result that indicates the existence of capacity achieving linear codes with the same low density of ones in each row of the generating matrix. In addition to proving the existence of capacity achieving sparse codes, we also show that for a sufficiently large code length, no search is necessary in practice to find the desired deterministic matrix. This means that any randomly chosen code can have the desired error correcting property with high probability. This is done by proving that the error probability of a sequence of codes, corresponding to a randomly selected sequence of sparse generating matrices tends to zero as $n$ approaches infinity, in probability. This important result is then extended to generating matrices with low density rows for the case of BSC.

Although in reality the bloclength of codes is finite, in order to prove that a class of codes is capacity achieving, we assume that the blocklength goes to infinity. An intersting question is that for a given error probability and blocklength, how close the rate of the code can be to the capacity. An upper bound for the channel coding rate achievable at a given blocklength and error probability is derived in \cite{Yuri}. In our paper we use Yuri's upper bound \cite{Yuri} and other well-known results to compare to our numerically derived results. 

An interesting trade-off between the sparsity of the generating matrix and the error exponent is demonstrated such that the sparser the matrix,  the smaller the error exponent becomes. It is important to note that for the case of BSC, we rigorously prove the existence of capacity achieving linear codes for a constant $\rho$ resulting in a non-vanishing density of ones on the generating matrix as $n$ tends to infinity. However, we have made a conjecture that if we choose $\rho(n)=1/n^\gamma$; where $0<\gamma<1$, the resulting codes can still be capacity achieving, which implies a vanishing density of ones. This signifies that the number of ones in the generating matrix can be as low as $n^{2-\gamma}$. For the case of BEC, we have been able to prove that to have capacity achieving generating matrices, $\rho(n)$ can be of $O(\frac{\log n}{n})$. This implies that the number of ones in the generating matrix is about $n\log n$ which is asymptotically less than $n^{2-\gamma}$, the number of ones in the case of BSC.  As opposed to the existing results in the literature, which are based on Maximum A Posteriori (MAP) decoders, the proposed proofs are based on a suboptimal decoder,\footnote{See the details in the next section.} which makes our approach also novel from decoder point of view.


The organization of the paper is as follows: In the next section, some preliminary definitions and notations are presented. In Sections \ref{sec:BSC} and \ref{sec:BEC}, we present our theorems for BSC and BEC, respectively, and Section \ref{sec:Con} concludes the paper.


\section{Preliminaries}\label{sec:Pre}
Consider a DMC which is characterized by $\mathcal{X}$ and $\mathcal{Y}$ as its input and output alphabet sets, respectively, and the transition probability function $\mathbb{P}(y|x)$, where $x\in\mathcal{X}$ is the input, and $y\in\mathcal{Y}$ is the output of the channel. In this paper, we consider the binary case where $\mathcal{X}=\{0,1\}$.
A binary code $\mathcal{C}(n,k)$ of rate $R$ is a mapping from the set of $2^{k}$ $k$-tuples $X_i$ to $n$-tuples $Z_i$, $0\le i\le 2^{k}-1$, where
$X_i\in\{0,1\}^k$, $Z_i\in\{0,1\}^n$, and the code rate $R$ is defined as the ratio of $k$ by $n$. Since we are only interested in \emph{Linear Codes},
the mapping is fully specified by an $n\times k$ binary matrix $\mathbf{A}=\{A_{ij}\}$ (the generating matrix), and encoding is accomplished by a left multiplication by $\mathbf{A}$:
\begin{align}
Z_i=\mathbf{A}X_i,\nonumber
\end{align}
where the calculations are in $\mathbb{GF}\left(2\right)$. The vector $Z_i$ is then transmitted through the DMC. Decoding is defined as recovering the vector $X_i$ from the possibly corrupted received version of $Z_i$.

In this paper the employed decoding scheme relies on the a posteriori probability distribution. Let $\mathbf{A}$ be the generating matrix. For a received vector $Y=y$, the decoder allocates a random vector such as $X=x$ as the original transmitted message with the conditional probability $\mathbb{P}(X=x|Y=y)$. Clearly, the probability of correct detection using $\mathbf{A}$ as the generating matrix is
\begin{equation}\label{eq:pc}
p_c(\mathbf{A})=\sum_{i,j}{\mathbb{P}(x_i)\mathbb{P}(y_j|x_i)\mathbb{P}(x_i|y_j)}=
\sum_{i,j}{\mathbb{P}(x_i,y_j)\mathbb{P}(x_i|y_j)}=\mathbb{E}_{X,Y}(\mathbb{P}(X|Y)),
\end{equation}
where $\mathbb{P}(X,Y)$  depends on $\mathbf{A}$.

Note that the optimal decoder is a MAP decoder which allocates $argmax_x{\mathbb{P}(X=x|Y=y)}$ and that the probability of correct detection using MAP is more than or equal to the probability of correct detection in (\ref{eq:pc}).
Throughout the paper, the index $i$ in $X_i$ and $Z_i$ may be dropped for more clarity. For the sake of convenience, the following notations are used for the remainder of the paper.
\begin{definition}
Let $\mathcal{A}_{n\times k}$ be the set of all binary $n\times k$ matrices. The density of an $\mathbf{A}\in\mathcal{A}_{n\times k}$ is defined as the total number of ones within the matrix divided by the number of its elements ($nk$). A matrix with a density less than $0.5$ is called sparse; the smaller the density, the sparser the matrix becomes.
\end{definition}
\begin{definition}
Let each entry of each element of $\mathcal{A}_{n\times k}$ has a Bernoulli($\rho$) distribution, $0<\rho <1$.\footnote{A binary random variable has Bernoulli($\rho$) distribution if it is equal to $1$ with probability of $\rho$ and equal to $0$ with probability of $1-\rho$.} This scheme induces a \emph{probability distribution} on the set $\mathcal{A}_{n\times k}$, denoted by Bernoulli($n,k,\rho$). For the rest of paper, we consider this distribution on the set $\mathcal{A}_{n\times k}$.
\end{definition}

Note that as $n$ approaches infinity, the typical matrices of $\mathcal{A}_{n\times k}$ have a density close to $\rho$.

\section{Binary Symmetric Channel (BSC)}\label{sec:BSC}
Consider a BSC with cross-over probability $\epsilon$. The capacity of this channel is given by $C=1-h(\epsilon)$, where $h\left(\epsilon\right)=-\epsilon\log\epsilon-\left(1-\epsilon\right)\log\left(1-\epsilon\right)$. We suppose that $R$, the rate of the code, is less than $C$. In this section, we prove the existence of capacity achieving linear codes with arbitrarily sparse generating matrices over the BSC. We prove the existence by showing that the average error probability over such generating matrices tends to zero as $n$ approaches infinity.
\subsection{Channel Modeling}
Assume that we encode a message vector $X$ to generate the codeword $\mathbf{A}X$. Note that $X$ is chosen uniformly from the set $\{0,1\}^k$. Due to the effect of error in the BSC, each entry of the transmitted codeword $\mathbf{A}X$ can be changed from $0$ to $1$ and vice versa. These changes can be modeled by adding $1$ to erroneous entries of $\mathbf{A}X$ (in $\mathbb{GF}(2)$). Therefore, the error of a BSC with cross-over probability $\epsilon$ can be modeled by a binary $n$-dimensional error vector $N$ with i.i.d. entries with Bernoulli($\epsilon$) distribution. Thus, if the output of the channel is shown by $Y$, the following equation models the channel:
\begin{eqnarray}\label{Eq:corretprob}
Y_{n\times 1}=\mathbf{A}_{n\times k}X_{k\times 1}+N_{n\times 1}.
\end{eqnarray}
Note that $X$ and $N$ are independent.

\subsection{Capacity achieving sparse linear codes for the BSC}
In the following theorem, a lower bound for the average probability of correct detection over the set $\mathcal{A}_{n\times k}$, is obtained.

\begin{theorem}\label{Th:aveerror} Consider a BSC with cross-over probability $\epsilon$. A lower bound for the average probability of correct detection over all $n \times k$ generating matrices with Bernoulli$(n,k,\rho)$ distribution is given by
\begin{equation}\label{eq:lowpc1}
\mathbb{E}_{\mathbf{A}\in\mathcal{A}_{n\times k}}\left(p_c\left(\mathbf{A}\right)\right)\geq
\sum_{i=0}^{n}\binom{n}{i}\times\frac{2^n{\epsilon^{2i}}{{(1-\epsilon)}^{2(n-i)}}}
{\sum_{j=0}^{k}\binom{k}{j}(1-(1-2\epsilon)(1-2\rho)^j)^i(1+(1-2\epsilon)(1-2\rho)^j)^{n-i}}.
\end{equation}
\end{theorem}
\emph{Proof}: See Appendix II.

\begin{note}
An important result of this theorem is that we can fix the error probability and find the maximal achievable rate for a given blocklength. See the following figures.
\end{note}

\begin{figure}\label{fig00}
\centering
\includegraphics[width=10cm]{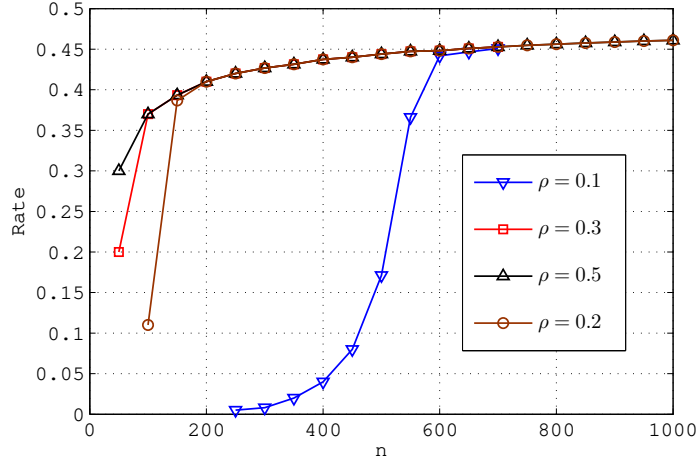}
\caption{A comparison of the coding rate versus the blocklength for various values of density($\rho$), $\epsilon=0.11$, Capacity=0.5, error probability=$10^{-1}$.}.
\end{figure}

Figure $1$ is a plot of the coding rate versus $n$ for $\rho$ equal to $0.1$, $0.3$ and $0.5$. This plot is numerically evaluated from Theorem \ref{Th:aveerror}. An interesting observation of this figure is that when the blocklength $n$ increases, the coding rate becomes independent of the density $\rho$. This observation can be shown to be true from (\ref{Eq:approx1term}) of Lemma \ref{lem:asymptotic}, where the parameter $\rho$ disappears on the right hand side. The significance of this observation is that sparse generating matrices can replace non-sparse ones for large block coding sizes, which implies simpler encoder design. This observation is the dual of LDPC codes where large sparse parity check matrices simplifies the decoder design, while the performance remains the same.

\begin{figure}\label{fig0}
\centering
\includegraphics[width=10cm]{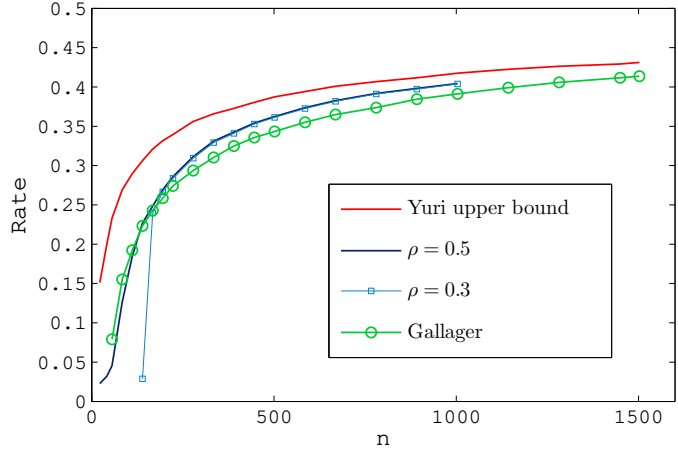}
\caption{A comparison of the coding rate versus the blocklength for various methods; the Gallager and the Yuri curves are plotted from \cite{Yuri}, $\epsilon=0.11$, Capacity=0.5, error probability=$10^{-3}$.}.
\end{figure}

Figure $2$ is a comparison of our result to that of Gallager result and Yuri upper bound \cite{Yuri}. This figure shows that our results are within the Yuri upper bound and the Gallager result. This figure also shows that for the probability of error equal to $10^{-3}$ when $n$ becomes greater than $180$, the performance of the sparse genearting matrices with $\rho=0.3$ becomes the same as the non-sparse matrices with $\rho=0.5$.
\\In the following theorem, we will show that the expected value of the correct detection probability over all generating matrices from $\mathcal{A}_{n\times k}$ approaches $1$. This proves the existence of at least one linear capacity achieving code.
\\
\begin{theorem}\label{Th:BSCgeneral} For any $0<\rho<1$, for a BSC we have
\begin{equation}
\lim_{n\to\infty}\mathbb{E}_{\mathbf{A}\in\mathcal{A}_{n\times k}}(p_c(\mathbf{A}))=1.
\end{equation}
\end{theorem}

\emph{Proof:} See Lemmas 1 and 2 and the proof in Appendix III.


The performance of linear codes is determined by the error exponent which is defined as follows:
\begin{definition}
The error exponent of a family of codes $\mathcal{C}$ of rate $R$ is defined as
\begin{eqnarray}\label{Eq:errorexponent}
E_{\mathcal{C}}(R)=\lim_{n\to\infty}-\frac{1}{n}\log p_e,
\end{eqnarray}
where $p_e$ is the average probability of decoding error.
\end{definition}

\begin{figure}\label{fig1}
\centering
\includegraphics[width=10cm]{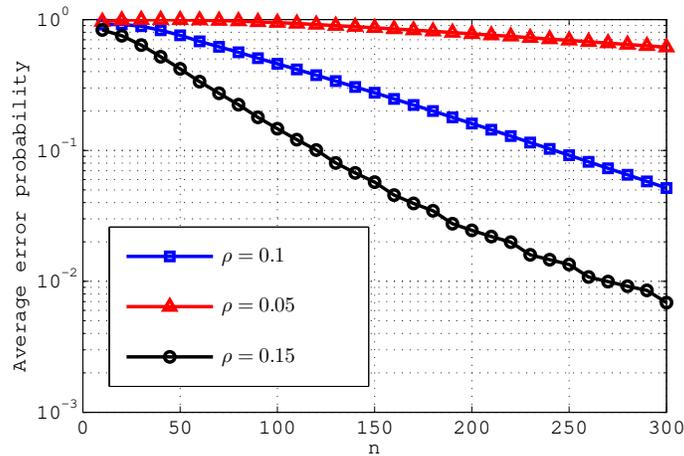}
\caption{The average error probability versus n, $\epsilon=0.05$, $R=0.8 C$}.
\end{figure}

If the limit is greater than zero, the average error probability of the proposed codes decreases exponentially to zero as $n$ increases. The error exponent is an index such that the larger the error exponent, the faster the probability of error decays as $n$ increases. Based on our observation, there is an interesting relation between the error exponent of the codes constructed by generating matrices with Bernoulli$(n,k,\rho)$ distribution and the values of $\rho$. In Fig. 3, we have plotted the average probability of error versus $n$ for various values of $\rho$. As it can be seen, the error exponent which is equal to the slope of the curves, increases as $\rho$ increases (the generating matrix become less sparse). In other words, although the probability of error for sparse codes goes to to zero exponentially as $n$ increases; this decrease is not as fast as high density codes.

\begin{definition}
\label{D6}
Let $W(A)$ be the number of ones in a given binary matrix $A$ and $\eta$ be an arbitrary positive constant. ${\mathcal{T}}^{\eta}_{n\times k}$ is defined as a subset of $\mathcal{A}_{n\times k}$ for which $|\frac{W(A)}{nk}-\rho|<\eta,\,\eta>0$. By choosing a sufficiently small $\eta$, the set ${\mathcal{T}}^{\eta}_{n\times k}$ is in fact a subset of ${\mathcal{A}}_{n\times k}$ which contains matrices having density of ones arbitrarily close to any given $\rho$. Note that the probability distribution on ${\mathcal{T}}^{\eta}_{n\times k}$ is induced from the probability distribution on ${\mathcal{A}}_{n\times k}$.
\end{definition}

In Theorems \ref{Th:aveerror} and \ref{Th:BSCgeneral}, we proved the existence of capacity achieving codes for any value of $\rho$. We did not explicitly prove the existence of sparse capacity achieving codes. However, using concentration theory \cite{concentration}, we can see that for a sufficiently large $n$, a randomly chosen matrix from ${\mathcal{A}}_{n\times k}$ is in the subset ${\mathcal{T}}^{\eta}_{n\times k}$ with high probability. In other words, we can state the following proposition which implies the existence of capacity achieving codes which are sparse.

\begin{proposition}\label{cor:typicalcol}
Let ${\mathcal{T}}^{\eta}_{n\times k}$ be the set of typical matrices defined in Definition (\ref{D6}). We then have
\begin{equation}
\lim_{n\to\infty}\mathbb{E}_{\mathbf{A}\in\mathcal{T}^{\eta}_{n\times k}}(p_e)=0.
\end{equation}
\end{proposition}

\begin{definition}
We define $\mathcal{R}_{n\times k}$ as the set of all binary $n\times k$ matrices with rows that have $k\rho$ ones. We also consider a uniform distribution on the set $\mathcal{R}_{n\times k}$ for the rest of the paper.
\end{definition}

In the next theorem, we will prove a stronger result on capacity achieving sparse codes. We show the existence of capacity achieving matrices with rows containing exactly $k\rho$ ones. In other words, the density of ones in each row is exactly equal to $\rho$. This also implies that the generating matrix has a density of ones exactly equal to $\rho$. In Theorem \ref{Th:aveerrorrow}, we shall derive a lower bound on the average probability of correct detection and in Theorem \ref{Th:BSCgenerrow} we will prove that this lower bound tends to one. This shows that the average probability of error over the set $\mathcal{R}_{n\times k}$ approaches zero, implying the existence of capacity achieving codes with generating matrices taken from $\mathcal{R}_{n\times k}$.

\begin{theorem}\label{Th:aveerrorrow} For a binary symmetric channel with cross-over probability $\epsilon$, a lower bound for the expected value of the probability of correct detection over all generating matrices in ${\mathcal{R}}_{n\times k}$ is given by
\begin{equation}\label{eq:lowpcrow1}
\mathbb{E}_{\mathbf{A}\in\mathcal{R}_{n\times k}}\left(p_c\left(\mathbf{A}\right)\right)\geq
\sum_{i=0}^{n}\binom{n}{i}{\epsilon^{i}}{{(1-\epsilon)}^{(n-i)}}\frac{{\epsilon^{i}}{{(1-\epsilon)}^{(n-i)}}}
{\sum_{j=0}^{k}\binom{k}{j}(\epsilon A_j+(1-\epsilon) B_j)^i((1-\epsilon) A_j +\epsilon B_j)^{n-i}}.
\end{equation}
where
\begin{equation}
A_j=\sum_{\text{q~is~odd}}\frac{1}{\binom{k}{k\rho}}\binom{j}{q}\binom{k-j}{k\rho-q},~~~
B_j=\sum_{\text{q~is~even}}\frac{1}{\binom{k}{k\rho}}\binom{j}{q}\binom{k-j}{k\rho-q}.\nonumber
\end{equation}
\end{theorem}
\emph{Proof}: See Appendix IV.
\begin{theorem}\label{Th:BSCgenerrow}
For each $0<\rho<1$, we have
\begin{equation}
\lim_{n\to\infty}\mathbb{E}_{\mathbf{A}\in\mathcal{R}_{n\times k}}(p_c(\mathbf{A}))=1.
\end{equation}
\end{theorem}
\emph{Proof:} See Lemma 3 and the proof in Appendix IV.

In Theorems \ref{Th:aveerror} and \ref{Th:BSCgeneral}, we proved the existence of capacity achieving linear codes with generating matrices having Bernoulli$(n,k,\rho)$ distribution by showing that the average probability of error over all generating matrices tends to zero as $n$ approaches infinity. This implies that we may have to perform a search over $\mathcal{A}_{n\times k}$ to find such a matrix. Assume that we simply pick matrices randomly for each $n$ from the set $\mathcal{A}_{n\times k}$. This constitutes a sequence of $n\times nR$ matrices.  Now consider the resulting sequence of error probabilities corresponding to the sequence of generating matrices. In the following proposition, we shall prove that the limit of this sequence is zero in probability, i.e., a sequence of randomly chosen matrices is capacity achieving with high probability. This suggests that for sufficiently large $n$, no search is necessary to find a desired deterministic generating matrix.
\begin{proposition}\label{cor:coninpro}
Let $\{\mathbf{A}_{n\times nR}\}_{n=0}^{\infty}$ be the sequence of matrices, where $\mathbf{A}_{n\times nR}$ is selected randomly from $\mathcal{A}_{n\times nR}$. If we denote the error probability of the generating matrix $\mathbf{A}_{n\times nR}$ over BSC by $p_e(\mathbf{A}_n)$, then $p_e(\mathbf{A}_n)$ converges in probability to zero as $n$ tends to infinity.
\end{proposition}
\emph{Proof:} See Appendix V.
\begin{note}
If we use the result of Theorem \ref{Th:BSCgenerrow}, we can extend Proposition  \ref{cor:coninpro} to the case where we construct the matrix sequence by choosing the matrices from the set $\mathcal{R}_{n\times k}$. In other words, in order to have capacity achieving sequences of generating matricescfor BSC with  arbitrarily low density rows, we can simply pick generating matrices randomly from $\mathcal{R}_{n\times k}$.
\end{note}

At this stage, we have been able to rigorously prove the existence of capacity achieving sparse linear codes over the BSC. However for a given $\rho$, although the density of ones can be made arbitrarily small, it does not go to zero even when $n$ approaches infinity. Let us assume the case where $\rho$ is a decreasing function of $n$ such that $\lim_{n\rightarrow \infty}\rho(n)=0$, resulting in zero density of ones as $n$ goes to infinity. In the following conjecture, we will propose a result indicating that this assumption can in fact be true. Although, we have not been able to rigorously prove the conjecture, a sketch of the proof has been presented in the appendix.


\begin{conjecture}\label{con:BSCgengen}
Let $\gamma$ be an arbitrary number from interval $(0,1)$. For $\rho(n)=\frac{1}{n^{\gamma}}$ by assuming the Bernoulli$(n,k,\rho(n))$ distribution on the set $\mathcal{A}_{n\times k}$, we have
\begin{equation}
\lim_{n\to\infty}\mathbb{E}_{\mathbf{A}\in\mathcal{A}_{n\times k}}(p_c(\mathbf{A}))=1
\end{equation}
\end{conjecture}
See Appendix V for the sketch of the proof.

\section{Binary Erasure Channel}\label{sec:BEC}
A binary erasure channel is identified by erasure probability $\epsilon$ and the capacity of this channel is given by $1-\epsilon$.
We use the decoder proposed in Section II. Through the channel, some entries of the coded vector $\mathbf{A}X$, shown by $Z$, may be erased. According to the position of the erased entries, the error of the channel can be modeled as a subset $F$ of $\mathcal{F}=\{1,\ldots,n\}$. Therefore, we employ a decoder which decides about the transmitted vector by observing only the non-erased entries denoted by $Z_F$. For each $i\in F$, the $i^{th}$ row of $\mathbf{A}$ is removed to derive $\mathbf{A}_F$. Therefore, the encoding and channel operation can be written as $Z_F={\mathbf{A}_F}X$. The decoder chooses $\hat{X}$, the estimation of $X$, randomly from the set $\mathcal{X}(Z,F)=\{X|{\mathbf{A}_F}X=Z_F\}$. In this case, the decoder is equivalent to the MAP decoder. From linear algebra, it can be shown that $|\mathcal{X}(Z,F)|=2^{k-rank(\mathbf{A}_F)}$, where $rank$ is the maximum number of independent rows of a matrix calculated in $\mathbb{GF}(2)$. Since $\hat{X}$ is chosen uniformly from $\mathcal{X}(Z,F)$, the probability of the correct detection of $X$ is equal to $2^{-\left(k-rank(\mathbf{A}_F)\right)}$. Thus, we have $p_{c|X,F}(\mathbf{A})=\mathbb{P}(\hat{X}=X|X,F)=2^{rank(\mathbf{A}_F)-k}$,
where $p_{c|X,F}$ represents the probability of correct detection when $X$ is transmitted and the position of erased entries are given in $F$.

\begin{theorem}\label{BECgeneral} Let $C$ be the capacity of a BEC and $\mathbf{A}\in\mathcal{A}_{n\times k}$ is a generating matrix corresponding to a code of rate $R<C$. For any $\rho(n)$ of $O(\frac{\log n}{n})$, the expected value of $p_c(\mathbf{A})$ over all matrices with Bernoulli$(n, k, \rho(n))$ distribution tends to $1$ as $n$ approaches infinity.
\begin{equation}
\lim_{n\to\infty}\mathbb{E}_{\mathbf{A}\in\mathcal{A}_{n\times k}}(p_c(\mathbf{A}))=1.
\end{equation}
\end{theorem}
\emph{Proof}: See Appendix VI.

From the concentration theory \cite{concentration}, similar to the case of the BSC, we can state the following proposition.
\begin{proposition}
For a BEC with capacity $C$, codes of rate $R<C$ and generating matrix from $\mathcal{T}^{\eta}_{n\times k}$, we have:
\begin{equation}
\lim_{n\to\infty}\mathbb{E}_{\mathbf{A}\in\mathcal{T}^{\eta}_{n\times k}}(p_e(\mathbf{A}))=0.
\end{equation}
\end{proposition}

\begin{figure}\label{fig2}
\centering
\includegraphics[width=10cm]{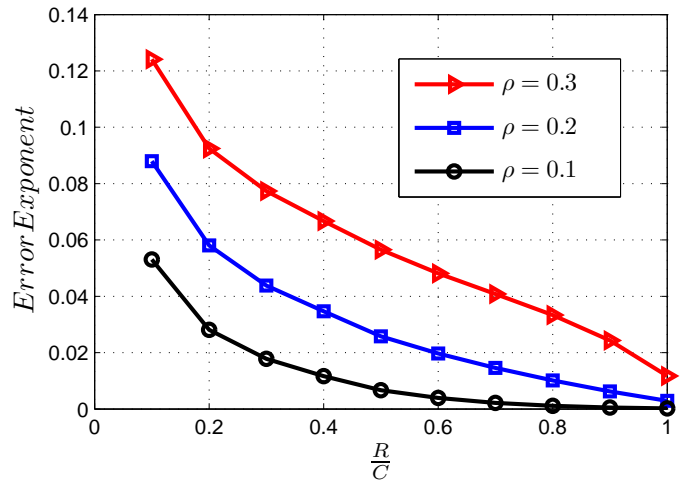}
\caption{The error exponent versus $\frac{R}{C}$ for different values of $\rho$ for BEC, $\epsilon=0.01$}.
\end{figure}

In the following proposition we show that similar to Proposition \ref{cor:coninpro} for BSC, a sequence of randomly chosen generating matrices from $\mathcal{A}_{n\times nR}$, results in a capacity achieving coding scheme with high probability. This suggests that for sufficiently large $n$, no search is necessary to find a desired deterministic generating matrix.

\begin{proposition}\label{cor:coninpro2}
Let $\{\mathbf{A}_{n\times nR}\}_{n=0}^{\infty}$ be the sequence of matrices, where $\mathbf{A}_{n\times nR}$ is selected randomly from $\mathcal{A}_{n\times nR}$. If we denote the error probability of the generating matrix $\mathbf{A}_{n\times nR}$ over BEC by $p_e(\mathbf{A}_n)$, then $p_e(\mathbf{A}_n)$ converges in probability to zero as $n$ tends to infinity.
\end{proposition}
\emph{Proof:} The proof is similar to the proof of Proposition \ref{cor:coninpro} and thus omitted.

In Fig. 2, we have shown the error exponent as a function of $\frac{R}{C}$ for different values of $\rho$. As it can be seen, a similar trade-off to BSC exists between sparsity and the error exponent. The smaller $\rho$ results in a smaller error exponent.

The following theorem is similar to Theorem \ref{Th:BSCgenerrow}.
\begin{theorem}\label{Th:BECrow} For each $0<\rho<1$, for a BEC we have
\begin{equation}
\lim_{n\to\infty}\mathbb{E}_{\mathbf{A}\in\mathcal{R}_{n\times k}}(p_c(\mathbf{A}))=1.
\end{equation}
\end{theorem}
\emph{Proof}: See Appendix VII.

\begin{note}
If we use the result of Theorem \ref{Th:BECrow}, we can extend Proposition  \ref{cor:coninpro2} to the case where we construct the matrix sequence by choosing the matrices from the set $\mathcal{R}_{n\times k}$.
\end{note}


\section{Conclusions}\label{sec:Con}
In this paper, a novel approach to prove the existence of capacity achieving sparse linear codes over the BSC and BEC was proposed. For the BSC, in Theorem \ref{Th:aveerror}, we derived a lower bound on the average probability of correct detection over the set $\mathcal{A}_{n\times k}$. In Theorem \ref{Th:BSCgeneral}, we proved that the average probability of error over $\mathcal{A}_{n\times k}$ tends to zero. Then we proved the existence of sparse capacity achieving codes in Proposition 2. In Theorem \ref{Th:aveerrorrow}, we derived a lower bound on the average probability of correct detection over the set $\mathcal{R}_{n\times k}$. Using this lower bound in Theorem \ref{Th:BSCgenerrow}, we proved the existence of capacity achieving codes with generating matrices with the same density $(\rho)$ in each row. In Proposition \ref{cor:coninpro} and its preceding note, we showed that the error probability of codes corresponding to any randomly chosen sequence of generating matrices tends to zero in probability. This implies that for sufficiently large $n$, a randomly chosen matrix from $\mathcal{A}_{n\times k}$ and $\mathcal{R}_{n\times k}$ will have the average error correcting capability. In addition, we conjectured that Theorem \ref{Th:BSCgeneral} can hold for the case where $\rho$ is a function of $n$, i.e. $\rho=1/n^\gamma$. This implies that for a capacity achieving code over a BSC, the density of the generating matrix can approach zero. In Theorem \ref{BECgeneral} and Proposition 3, we proved the existence of sparse codes for the case of BEC with generating matrices having Bernoulli distribution with $\rho(n)$ of $O(\frac{\log n}{n})$. A similar result to Proposition \ref{cor:coninpro} and Theorem \ref{Th:BSCgenerrow} was shown for BEC in Proposition \ref{cor:coninpro2} and Theorem \ref{Th:BECrow}, respectively. We demonstrated an interesting trade-off between the sparsity of the generating matrix and the error exponent indicating that a sparser generating matrix results in a smaller error exponent. We also observed that for small block sizes, generating matrices with higher densities have better performances while for large block sizes, the performance of the random generating matrices become independent of the density. In our proofs, we have used a suboptimal decoder while previous works in the literature were based on a MAP decoder. This implies that we can get stronger results if we use the optimal MAP decoder.

For future work, one can try to rigorously prove Conjecture 1 and possibly extend it to the case of matrices in the set $\mathcal{R}_{n\times k}$. The improvement in the bounds using a MAP decoder can be an interesting topic to investigate. The extension of the results to DMC's is another challenging topic to be explored. A very interesting work is to analytically derive the error exponent to  prove the trade-off between error exponent and sparsity of the generating matrix.

\newpage
\appendices
\section{\textbf{}}
We need the following definitions in order to prove our theorems.
\begin{definition}
Any two functions $a(n)$ and $b(n)$ are referred to as \emph{proportionally  equivalent } and written as $a(n)\approx b(n)$ if $
\lim_{n\to\infty}\frac{a(n)}{b(n)}=1$.
\end{definition}
\begin{definition}
Any two functions $c(n)$ and $d(n)$ are referred to as \emph{differentially equivalent} and written as $c(n)\doteq d(n)$ if $\lim_{n\to\infty}{c(n)-d(n)}=0$.
\end{definition}
\section{\textbf{The Proof of Theorem \ref{Th:aveerror}}}
\textbf{\emph{Proof of Theorem \ref{Th:aveerror}}}:
According to (\ref{eq:pc}), Bayes' rule, and the independency of $X$ and $N$, we have
\begin{eqnarray}\label{Eq:probcor}
p_c(\mathbf{A})=\mathbb{E}_{X,Y}\left(\frac{\mathbb{P}(X)\mathbb{P}(Y|X)}{\mathbb{P}(Y)}\right)=
\mathbb{E}_{X,N}\left(\frac{\mathbb{P}(N)\mathbb{P}(X)}{\mathbb{P}(Y)}\right).
\end{eqnarray}
Taking expectation over all matrices $\mathbf{A}\in \mathcal{A}_{n\times k}$, we get
\begin{align}
\mathbb{E}_{\mathbf{A}}(p_c(\mathbf{A}))
=\mathbb{E}_{\mathbf{A},X,N}\left(\frac{\mathbb{P}(N)\mathbb{P}(X)}{\mathbb{P}(Y)}\right)
=\mathbb{E}_{X,N}\left(\mathbb{P}(N)\mathbb{P}(X)
\mathbb{E}_{\mathbf{A}}\left(\frac{1}{\mathbb{P}(Y)}\right)\right),
\end{align}
where in the last equality, the independency among $\mathbf{A}$, $N$ and $X$ is used. Using the Jensen's inequality (see \cite{cover}, Chapter 2, Page 25), we have
\begin{align}\label{eq:lowpc}
\mathbb{E}_{\mathbf{A}}(p_c(\mathbf{A}))\ge&
\mathbb{E}_{X,N}\left(\frac{\mathbb{P}(N)\mathbb{P}(X)}{\mathbb{E}_{\mathbf{A}}(\mathbb{P}(Y))}\right)
\nonumber\\=&\mathbb{E}_{X,N}\left(\frac{\mathbb{P}(N)\mathbb{P}(X)}{\mathbb{E}_{\mathbf{A}}(\mathbb{P}_{X',N'}(\mathbf{A}X+N=\mathbf{A}X'+N'))}\right)
\nonumber\\
=&\mathbb{E}_{X,N}\left(\frac{\mathbb{P}(N)\mathbb{P}(X)}{\mathbb{E}_{\mathbf{A}}\left(\mathbb{E}_{X',N'}(1_{[\mathbf{A}(X-X')+(N-N')=0]})\right)}\right)\nonumber
\nonumber\\
=&\mathbb{E}_{X,N}\left(\frac{\mathbb{P}(N)\mathbb{P}(X)}{\mathbb{E}_{X'}(\mathbb{P}_{\mathbf{A},N'}(\mathbf{A}(X-X')+(N-N')=0))}\right),
\end{align}
where ${X'}$ and $N'$ have the same distributions as the input and error vectors, respectively. In the above equation, the expected value over $X'$ is a function of binary subtraction $X-X'$ and as a result does not depend on $X$. Thus we can assume any binary vector X such as the all zero vector, $X_0$; from the independency of the rows of $\mathbf{A}$ in (\ref{eq:lowpc}) and the uniformity of the vectors $X$, we have

\begin{align}\label{Eq:prooflowbound}
\mathbb{E}_{\mathbf{A}}(p_c(\mathbf{A}))\ge&\frac{1}{2^k}\mathbb{E}_{N,X=
X_0}\left(\frac{\mathbb{P}(N)}
{\mathbb{E}_{X'}\left(\prod_{l=1}^{n}\mathbb{P}_{A_l,N'_l}\left(A_l(X_0-X')+(N_l-N'_l)=0\right)\right)}\right)\\\nonumber
&=\frac{1}{2^k}\mathbb{E}_{N}\left(\frac{\mathbb{P}(N)}
{\mathbb{E}_{X'}\left(\prod_{l=1}^{n}\mathbb{P}_{A_l,N'_l}\left(A_l(X')+(N_l-N'_l)=0\right)\right)}\right),
\end{align}
where $N_l$ and ${N_l}'$ are the $l^{th}$ entry of $N$ and $N'$, respectively, and $A_l$ is the $l^{th}$ row of $\mathbf{A}$. Note that here all the operations are performed in $\mathbb{GF}{(2)}$. In order to evaluate the right side of the above inequality, assume that vector $N$ has $i$ ones. Without loss of generality and for convenience, we assume that the first $i$ elements of $N$ are $1$. Thus, the argument of the expected value in (\ref{Eq:prooflowbound}) is equal to
\begin{equation}\label{eq16}
\frac{\epsilon^i (1-\epsilon)^{n-i}}{\mathbb{E}_{X'}\left(\prod_{l=1}^{i}\left(\mathbb{P}_{A_l,N'_l}\left(A_lX'+N'_l=1\right)\right)
\prod_{l=i+1}^{n}\left(\mathbb{P}_{A_l,N'_l}\left(A_lX'+N'_l=0\right)\right)\right)}.
\end{equation}

To evaluate the expected value in the above expression, note that since $N'_l=0$ with probability $1-\epsilon$ and $N'_l=1$ with probability $\epsilon$, we have
\begin{align} \label{Eq:diff-proof}
\prod_{l=1}^{i}\left(\mathbb{P}_{A_l,N'_l}\left(A_lX'+N'_l=1\right)\right)=
\left(\epsilon\times\mathbb{P}\left(A_lX'=0\right)+(1-\epsilon)\times\mathbb{P}
\left(A_lX'=1\right)\right)^i.
\end{align}
Now assume $j$ elements of $X'$ are equal to $1$. Also consider the entries of $A_l$ with the same indices as the entries of $X'$ that are equal to one. It is easy to see that in the above equation, $\mathbb{P}\left(A_lX'=1\right)$ is equal to the probability of having an odd number of ones in the considered indices of $A_l$. Thus, we have
\begin{align}\label{Eq:difwith}
\mathbb{P}\left(A_lX'=1\right)=
\sum_{\text{q~odd}}\binom{j}{q}\rho^q(1-\rho)^{j-q}=
\frac{((1-\rho)+\rho)^j-((1-\rho)-\rho)^j}{2})=\frac{1-(1-2\rho)^j}{2}.
\end{align}
The same argument results in $\mathbb{P}\left(A_lX'=0\right)=\frac{1+(1-2\rho)^j}{2}$ and therefore we have
\begin{align}\label{eq19}
\prod_{l=1}^{n}\mathbb{P}_{A_l,N'_l}\left(A_l(X_0-X')+(N_l-N'_l)=0\right)
=\left(\frac{1-(1-2\epsilon)(1-2\rho)^j}{2}\right)^i\left(\frac{1+(1-2\epsilon)
(1-2\rho)^j}{2}\right)^{n-i}.
\end{align}
The expectation of the above expression over $X'$ results in
\begin{align}\label{Eq:prooftemp}
&\mathbb{E}_{X'}\left(\prod_{l=1}^{n}\mathbb{P}_{A_l,N'_l}\left(A_l(X_0-X')+(N_l-N'_l)=0\right)\right)
=\nonumber\\&\sum_{j=0}^{k}\frac{1}{2^k}\binom{k}{j}\left(\frac{1-(1-2\epsilon)(1-2\rho)^j}{2}\right)^i
\left(\frac{1+(1-2\epsilon)(1-2\rho)^j}{2}\right)^{n-i}.
\end{align}
Substituting (\ref{Eq:prooftemp}) in (\ref{eq16}) and taking expected value with respect to $N$, we obtain the following lower bound for $\mathbb{E}_{\mathbf{A}}(p_c)$:
\begin{align}
\mathbb{E}_{\mathbf{A}\in\mathcal{A}_{n\times k}}(p_c(\mathbf{A}))\ge
\sum_{i=0}^{n}\binom{n}{i}
{\epsilon}^i{(1-\epsilon)}^{n-i}
\frac{2^n{\epsilon}^i{(1-\epsilon)}^{n-i}}
{\sum_{j=0}^{k}(1-(1-2\epsilon)(1-2\rho)^j)^i(1+(1-2\epsilon)(1-2\rho)^j)^{n-i}}.\nonumber
\end{align}
This completes the proof. $\blacksquare$
\section{\textbf{Lemmas \ref{cramergen}, \ref{lem:asymptotic} and the proof of Theorem \ref{Th:BSCgeneral}}}
\begin{lemma}\label{cramergen}
Let $\{a_i\}_{i=0}^{\infty}$ be a bounded sequence. For any $\delta> 0$ and $0\le p\le 1$ the summation $\sum_{i=0}^{n}\binom{n}{i}p^i(1-p)^{n-i}a_i$ is differentially equivalent to $\sum_{i=n(p-\delta)}^{n(p+\delta)}\binom{n}{i}p^i(1-p)^{n-i}a_i$.
\end{lemma}
\emph{Proof:} According to the Chernoff-Hoeffding Theorem \cite{chernof} the proof is straightforward. $\blacksquare$
\\\\
\begin{lemma}\label{lem:asymptotic} Consider a code with rate $R$ over a BSC with cross-over probability of $\epsilon$ where $R=k/n<C=1-h(\epsilon)$. There exists a $\delta>0$ for which for any $i\in(n(\epsilon-\delta),n(\epsilon+\delta))$, we have
\begin{align}\label{Eq:approx1term}
\sum_{j=0}^{k}\binom{k}{j}(1-(1-2\epsilon)(1-2\rho)^j)^i(1+(1-2\epsilon)(1-2\rho)^j)^{n-i}\approx 2^n \epsilon^i(1-\epsilon)^{n-i}.
\end{align}
\end{lemma}
\emph{Proof:} 
To prove the lemma, note that the first term of the summation $(j=0)$ in the left hand side of (\ref{Eq:approx1term}) is equal to the right hand side. Therefore, to prove (\ref{Eq:approx1term}), it is sufficient to show that
\begin{equation}\label{eqmid}
\sum_{j=1}^{k}\binom{k}{j}\left(\frac{1-(1-2\epsilon)(1-2\rho)^j}{2\epsilon}\right)^{i}\left
(\frac{1+(1-2\epsilon)(1-2\rho)^j}{2(1-\epsilon)}\right)^{n-i}  \doteq 0 .
\end{equation}
Let
$b(j)=\frac{1-(1-2\epsilon)(1-2\rho)^j}{2}$ and
$a(j,i)=\left(\left(\frac{1-(1-2\epsilon)(1-2\rho)^j}{2\epsilon}\right)^{\frac{i}{n}}
\left(\frac{1+(1-2\epsilon)(1-2\rho)^j}{2(1-\epsilon)}\right)^{\frac{n-i}{n}}\right)^{\frac{1}{R}}$. Thus, we have
\begin{equation}
a(j,i)=\left(\left(\frac{b(j)}{\epsilon}\right)^{\frac{i}{n}}\left(\frac{1-b(j)}{1-\epsilon}\right)^{\frac{n-i}{n}}\right)^{\frac{1}{R}}.
\end{equation}
By using a straightforward calculation, it can be shown that for $i=n\epsilon$, the maximum of $a(j,i)$ is equal to $1$. The maximum of $a(j,i)$ occurs for $b(j)=\epsilon$ or equivalently $j=0$. Thus, for $j\ge 1$ we have
\begin{equation}\label{Eq:fiinep}
a(j,n\epsilon) < 1.
\end{equation}
Also, since $\lim_{j\to\infty}(b(j))=\frac{1}{2}$, we have
\begin{equation}\label{Eq:inep}
\lim_{j\to\infty} a(j,n\epsilon)=\left(\left(\frac{1}{2\epsilon}\right)^{\epsilon}\left(\frac{1}{2(1-\epsilon)}\right)^{1-\epsilon}\right)^{\frac{1}{R}}
=2^{-\frac{C}{R}}<\frac{1}{2}.
\end{equation}
It is easy to see that $a(j,i)$ is a uniformly continuous function of $i$ and $j$. Thus from (\ref{Eq:fiinep}), we conclude that there is a ${\delta}_1>0$ for which for any $i\in(n(\epsilon-{\delta}_1),n(\epsilon+{\delta}_1))$ and $j\ge 1$, we have $a(j,i) < 1$. And also from (\ref{Eq:inep}), we conclude that there is a ${\delta}_2>0$ for which for any $i\in(n(\epsilon-{\delta}_2),n(\epsilon+{\delta}_2))$, we have $\lim_{j\to\infty} a(j,i)<\frac{1}{2}$. Let $\delta=\min({\delta}_1,{\delta}_2)$ and fix $i\in(n(\epsilon-{\delta}),n(\epsilon+{\delta}))$; there exist an integer $M$ and a real number $\mu>0$, for which we have $a(j,i)<\frac{1}{2}-\mu$ for all $j>M$. By using this $M$, the left hand side of (\ref{eqmid}) can be written as
\begin{eqnarray}\label{eq26-1}
\sum_{j=1}^{k}\binom{k}{j}a(j,i)^k=\sum_{j=1}^{M}\binom{k}{j}a(j,i)^k+
\sum_{j=M+1}^{k}\binom{k}{j}a(j,i)^k.
\end{eqnarray}
Since $a(j,i)<\frac{1}{2}-\mu$ for $j>M$, we have
\begin{align}
&\lim_{k\to\infty}\sum_{j=M+1}^{k}\binom{k}{j}a(j,i)^k\leq\lim_{k\to\infty} \left(\sum_{j=M+1}^{k}\binom{k}{j}\right)(\frac{1}{2}-\mu)^k\leq\lim_{k\to\infty}2^k(\frac{1}{2}-\mu)^k=0.
\end{align}
Therefore, $\lim_{k\to\infty}\sum_{j=M+1}^{k}\binom{k}{j}a(j,i)^k=0$.\\
To see that the first term at the right hand side of (\ref{eq26-1}) also tends to zero, let $w=\max_{1\le j\le M}a(j,i)<1$. Therefore, we can write
\begin{align}
\sum_{j=1}^{M}\binom{k}{j}a(j,i)^k<\left(\sum_{j=1}^{M}\binom{k}{j}\right){w}^k\le
\left(Mk^M\right){w}^k=Me^{-{({\upsilon}k-M\ln(k)})},\nonumber
\end{align}
where $\upsilon=-\ln(w)>0$. Now the right hand side of the above inequality tends to zero because ${\upsilon}k-M\ln(k)$ tends to infinity as $k$ approaches infinity. This proves that the left hand side should also tend to zero. Therefore, both summations at the right hand side of (\ref{eq26-1}) tend to zero. This proves (\ref{eqmid}) and consequently (\ref{Eq:approx1term}). $\blacksquare$

\textbf{\emph{Proof of Theorem \ref{Th:BSCgeneral}}}:
\\
Let $a_i=\frac{2^n\epsilon^i(1-\epsilon)^{n-i}}
{\sum_{j=0}^{k}\binom{k}{j}(1-(1-2\epsilon)(1-2\rho)^j)^i(1+(1-2\epsilon)(1-2\rho)^j)^{n-i}}$. The first term of the summation of the denominator is equal to the numerator, and the other terms in the summation are positive. Thus, the elements of the sequence $\{a_i\}_{i=0}^{n}$ are less than $1$ and subsequently bounded. Therefore, we can apply Lemma 1. 
Now note that based on Theorem \ref{Th:aveerror}, we have
\begin{equation}
\mathbb{E}_{\mathbf{A}\in\mathcal{A}_{n\times k}}(p_c(\mathbf{A}))\ge \sum_{i=0}^{n}\binom{n}{i}{\epsilon}^i{(1-\epsilon)}^{n-i}
\frac{2^n{\epsilon}^i{(1-\epsilon)}^{n-i}}
{\sum_{j=0}^{k}(1-(1-2\epsilon)(1-2\rho)^j)^i(1+(1-2\epsilon)(1-2\rho)^j)^{n-i}}.\nonumber
\end{equation}
Let $\delta$ be as in Lemma 1. Since $\mathbb{E}_{\mathbf{A}\in\mathcal{A}_{n\times k}}(p_c(\mathbf{A}))\leq 1$, to prove the theorem, it is enough to show that the right hand side of the above inequality is differentially equivalent to $1$. To see this, we write
\begin{equation}
\sum_{i=0}^{n}\binom{n}{i}{\epsilon}^i{(1-\epsilon)}^{n-i}\frac{2^n{\epsilon}^i{(1-\epsilon)}^{n-i}}
{\sum_{j=0}^{k}(1-(1-2\epsilon)(1-2\rho)^j)^i(1+(1-2\epsilon)(1-2\rho)^j)^{n-i}}\doteq
\nonumber
\end{equation}
\begin{equation}
\sum_{i=n(\epsilon-\delta)}^{n(\epsilon+\delta)}\binom{n}{i}{\epsilon}^i{(1-\epsilon)}^{n-i}\frac{2^n{\epsilon}^i{(1-\epsilon)}^{n-i}}
{\sum_{j=0}^{k}(1-(1-2\epsilon)(1-2\rho)^j)^i(1+(1-2\epsilon)(1-2\rho)^j)^{n-i}}\doteq
\end{equation}
\begin{equation}
\sum_{i=n(\epsilon-\delta)}^{n(\epsilon+\delta)}\binom{n}{i}{\epsilon}^i{(1-\epsilon)}^{n-i}
\frac{2^n{\epsilon}^i{(1-\epsilon)}^{n-i}}{2^n{\epsilon}^i{(1-\epsilon)}^{n-i}}\nonumber\doteq
\sum_{i=0}^{n}\binom{n}{i}{\epsilon}^i{(1-\epsilon)}^{n-i}=1,\nonumber
\end{equation}
where we used Lemma 1 in the first and third equality\footnote{Note that by equality we mean $\doteq$ which is not mathematically precise but we use it throughout the paper for the ease of explanation.} and we replaced the summation in the denominator based on Lemma 2.
This proves the theorem.$\blacksquare$

\section{\textbf{The proof of Theorems \ref{Th:aveerrorrow} and \ref{Th:BSCgenerrow}}}
\textbf{\emph{Proof of Theorem \ref{Th:aveerrorrow}}}:
\\
We follow steps similar to that of Theorem \ref{Th:aveerror}. Equations (\ref{Eq:probcor}) to (\ref{eq16}) in Theorem \ref{Th:aveerror} still hold here.
It can be easily seen that
$\mathbb{P}\left(A_lX'=1\right)=A_j$ and $\mathbb{P}\left(A_lX'=0\right)=B_j$.  Thus equation (\ref{Eq:diff-proof}) is modified  as
\begin{align}
\prod_{l=1}^{n}\mathbb{P}_{A_l,N'_l}\left(A_l(X')+(N_l-N'_l)=0\right)
=\left(\epsilon A_j + (1-\epsilon) B_j\right)^i\left(\epsilon B_j + (1-\epsilon) A_j \right)^{n-i}.
\end{align}
The expectation of the above expression over $X'$ results in
\begin{align} \label{eq:proof3temp}
&\mathbb{E}_{X'}\left(\prod_{l=1}^{n}\mathbb{P}_{A_l,N'_l}\left(A_l(X_0-X')+(N_l-N'_l)=0\right)\right)
=\nonumber\\&\sum_{j=0}^{k}\frac{1}{2^k}\binom{k}{j}\left(\epsilon A_j + (1-\epsilon) B_j\right)^i\left(\epsilon B_j + (1-\epsilon) A_j \right)^{n-i}.
\end{align}
Substituting (\ref{eq:proof3temp}) into (\ref{eq16}) and taking the expectation with respect to $N$, we obtain
\begin{align}
\mathbb{E}_{\mathbf{A}\in\mathcal{R}_{n\times k}}(p_c(\mathbf{A}))\ge
\sum_{i=0}^{n}\binom{n}{i}
{\epsilon}^i{(1-\epsilon)}^{n-i}
\frac{{\epsilon}^i{(1-\epsilon)}^{n-i}}
{\sum_{j=0}^{k}\binom{k}{j}\left(\epsilon A_j + (1-\epsilon) B_j\right)^i\left(\epsilon B_j + (1-\epsilon) A_j \right)^{n-i}}.\nonumber
\end{align}
This completes the proof. $\blacksquare$

\textbf{\emph{Proof of Theorem \ref{Th:BSCgenerrow}}}:
\\
First we prove a lemma similar to Lemma \ref{lem:asymptotic}.
\begin{lemma}\label{lem:asymptoticrow}
Suppose that $R=k/n < 1-h(\epsilon)$. There exists a $\delta>0$ for which for any $i\in(n(\epsilon-\delta),n(\epsilon+\delta))$, we have
\begin{align}\label{eq:lemma3}
{\sum_{j=0}^{k}\binom{k}{j}\left(\epsilon A_j + (1-\epsilon) B_j\right)^i\left(\epsilon B_j + (1-\epsilon) A_j \right)^{n-i}}\approx \epsilon^i(1-\epsilon)^{n-i}.
\end{align}
\end{lemma}
\emph{Proof:}
Let $b(j)=\epsilon A_j + (1-\epsilon) B_j$ and $a(j,i)
=\left(\left(\frac{\epsilon A_j + (1-\epsilon) B_j}{\epsilon}\right)^{\frac{i}{n}}\left(\frac{\epsilon B_j + (1-\epsilon) A_j}{1-\epsilon} \right)^{\frac{n-i}{n}}\right)^{\frac{1}{R}}.$ Since $A_j+B_j=1$, we have
\begin{align}
a(j,i)
=\left(\left(\frac{b(j)}{\epsilon}\right)^{\frac{i}{n}}\left(\frac{1-b(j)}{1-\epsilon} \right)^{\frac{n-i}{n}}\right)^{\frac{1}{R}}.
\end{align}
By employing the same approach as the proof of Lemma \ref{lem:asymptotic}, it is sufficient to show that
 \begin{align}
\lim_{j\to\infty} b(j)=\frac{1}{2}.\nonumber
\end{align}

It is easy to see that $\lim_{j\to\infty}A_j=\lim_{j\to\infty}B_j=\frac{1}{2}$. As a result, we have
\begin{align}
&\lim_{j\to\infty} b(j)=\lim_{j\to\infty}{\left(\epsilon A_j + (1-\epsilon) B_j\right)}=\frac{\epsilon}{2}+\frac{1-\epsilon}{2}=\frac{1}{2}.
\end{align}
This completes the proof of lemma. $\blacksquare$

Now to prove this theorem, it is enough to replace the denominator in summation of (\ref{eq:lowpcrow1}) with the right hand side of (\ref{eq:lemma3}) according to Lemma \ref{lem:asymptoticrow}.
$\blacksquare$

\section{\textbf{Proof of Proposition \ref{cor:coninpro} and Proof Sketch of Conjecture \ref{con:BSCgengen}} }
\textbf{\emph{Proof of Proposition \ref{cor:coninpro}}}:
\\
 In order to show that $\lim_{n\to\infty}p_e(\mathbf{A}_{n\times nR})=0$ in probability, we have to show that for any given $\delta>0$,
\begin{equation}
\lim_{n\to\infty}\mathbb{P}(p_e(\mathbf{A}_{n\times nR})>\delta)=0.\nonumber
\end{equation}
For a given $\xi>0$, define $\beta=\min\{\delta,\xi\}$. According to Theorem \ref{Th:BSCgeneral}, we have $\lim_{n\to\infty}{\mathbb{E}(p_e(\mathbf{A}_{n\times nR})))}=0$. Thus, there exists an $N_\beta$ for which for any $n>N_\beta$, $\mathbb{E}(p_e(\mathbf{A}_{n\times nR}))<\beta^2$. Therefore, due to the fact that $p_e(\mathbf{A}_{n\times nR})\ge 0$, for $n>N_\beta$, we obtain $\mathbb{P}(p_e(\mathbf{A}_{n\times nR})>\beta)<\beta$. Hence, for $n>N_\beta$, since $\beta\le\delta$, we have
\begin{align}
&\mathbb{P}(p_e(\mathbf{A}_{n\times nR})>\delta)\le\mathbb{P}(p_e(\mathbf{A}_{n\times nR})>\beta)<\beta<\xi.\nonumber
\end{align}
Thus, for $n>N_\beta$, we have
\begin{equation}
\mathbb{P}(p_e(\mathbf{A}_{n\times nR})>\delta)<\xi,\nonumber
\end{equation}
and the proof is complete. $\blacksquare$

\textbf{\emph{Sketch of proof of Conjecture \ref{con:BSCgengen}}}:
\\
The lower bound of Theorem \ref{Th:aveerror} still holds for the case where $\rho$ is a function of $n$ where $0<\rho(n)<1$. If we can show that
for $R=k/n < 1-h(\epsilon)$, there exists a $\delta>0$ for which for any $i\in(n(\epsilon-\delta),n(\epsilon+\delta))$, we have
\begin{align}
\sum_{j=0}^{k}\binom{k}{j}\left(1-(1-2\epsilon)(1-2\rho(n))^j\right)^i\left(1+(1-2\epsilon)(1-2\rho(n))^j\right)^{n-i}\approx 2^n \epsilon^i(1-\epsilon)^{n-i}.
\end{align}

From the approach similar to that of the proof of Theorem \ref{Th:BSCgeneral}, the proof will be straightforward. Although, we have numerical evidence suggesting that the above equality holds, we have not been able to prove it rigorously. The rest of the proof is as follows. Let
\begin{equation}
 a_i=\frac{2^n\epsilon^i(1-\epsilon)^{n-i}}
{\sum_{j=0}^{k}\binom{k}{j}(1-(1-2\epsilon)(1-2\rho(n))^j)^i(1+(1-2\epsilon)(1-2\rho(n))^j)^{n-i}}.
\end{equation}
Since, the first term of the summation in the denominator is equal to the numerator, the sequence $\{a_i\}_{i=0}^{n}$ are less than $1$ and subsequently bounded. From Lemma \ref{cramergen} we get
\begin{eqnarray}
\sum_{i=0}^{n}\binom{n}{i}{\epsilon}^i{(1-\epsilon)}^{n-i}a_i\doteq\sum_{i=n(\epsilon-\delta)}
^{n(\epsilon+\delta)}\binom{n}{i}\epsilon^i(1-\epsilon)^{n-i}a_i.\nonumber
\end{eqnarray}
Now we have
\begin{align}
\mathbb{E}_\mathbf{A}(p_c(\mathbf{A}))\ge&\sum_{i=0}^{n}\binom{n}{i}{\epsilon}^i{(1-\epsilon)}^{n-i}
\frac{2^n{\epsilon}^i{(1-\epsilon)}^{n-i}}
{\sum_{j=0}^{k}(1-(1-2\epsilon)(1-2\rho(n))^j)^i(1+(1-2\epsilon)(1-2\rho(n))^j)^{n-i}}\nonumber\\\doteq&
\sum_{i=n(\epsilon-\delta)}^{n(\epsilon+\delta)}\binom{n}{i}{\epsilon}^i{(1-\epsilon)}^{n-i}
\frac{2^n{\epsilon}^i{(1-\epsilon)}^{n-i}}{2^n{\epsilon}^i{(1-\epsilon)}^{n-i}}\nonumber\\\doteq&
\sum_{i=0}^{n}\binom{n}{i}{\epsilon}^i{(1-\epsilon)}^{n-i}=1\nonumber.
\end{align}
In other words, $\lim_{n\to\infty}\mathbb{E}_{\mathbf{A}\in\mathcal{A}_{n\times k}}(p_c(\mathbf{A}))=1$ which is the desired result.$\blacksquare$

\section{\textbf{The Proof of Theorem \ref{BECgeneral}}}
\textbf{\emph{Proof of Theorem \ref{BECgeneral}}}:
\\
We first present a lemma from \cite{rank}.
\\
\begin{lemma}\label{lem:ref}
Suppose $\delta\ge 0$ and let Bernoulli$(n\left(1+\delta\right), n, \rho(n))$ be the probability distribution on the $n\left(1+\delta\right) \times n$ matrices where $\rho(n)$ is of $O(\frac{\log n} {n})$. Then $\mathbb{E}\left(rank\left(\mathbf{A}_{n(1+\delta) \times n}\right)\right)\approx n$.
\end{lemma}

Since $k/n<1-\epsilon$, it can be concluded that there exists a $\delta>0$ for which $k=n(1-\epsilon-\delta)$. By using the proposed decoding scheme and by decomposing $p_{c|X}(\mathbf{A})$ according to the position of the erased entries $F$, we get
\begin{align}
p_{c|X}(\mathbf{A})=\mathbb{P}(\hat{X}=X|X)=\sum_{F\subseteq\mathcal{F}}\mathbb{P}(\hat{X}=X|X,F)\mathbb{P}(F)
=\sum_{F\subseteq\mathcal{F}}{\epsilon}^{|F|}{(1-\epsilon)}^{n-{|F|}}2^{rank(\mathbf{A}_F)-k}.\nonumber
\end{align}
Therefore, $p_{c|X}(\mathbf{A})$ is the same for all $X$'s. Thus $p_c(\mathbf{A})=p_{c|X}(\mathbf{A})$. By evaluating the expected value of $p_c(\mathbf{A})$ over all matrices and using Jensen inequality, we have
\begin{eqnarray}
\mathbb{E}_{\mathbf{A}}{(p_c(\mathbf{A}))}=\sum_{i=0}^{n}\binom{n}{i}{\epsilon}^i
{(1-\epsilon)}^{n-i}\mathbb{E}_{\mathbf{A}}\left(2^{rank({\mathbf{A}}_{\left(n-i\right)\times k})-k}\right) 
\ge\sum_{i=0}^{n}\binom{n}{i}{\epsilon}^i
{(1-\epsilon)}^{n-i}2^{\mathbb{E}_{\mathbf{A}}\left(rank(\mathbf{A}_{\left(n-i\right)\times k})\right)-k}\nonumber.
\end{eqnarray}
Applying Lemma \ref{cramergen}, we obtain
\begin{align}
\sum_{i=0}^{n}\binom{n}{i}{\epsilon}^i{(1-\epsilon)}^{n-i}2^{\mathbb{E}_{\mathbf{A}}\left(rank(\mathbf{A}_{\left(n-i\right)\times k})\right)-k}\doteq\sum_{i=n(\epsilon-\theta)}^{n(\epsilon+\theta)}\binom{n}{i}{\epsilon}^i
{(1-\epsilon)}^{n-i}2^{\mathbb{E}_{\mathbf{A}}\left(rank(\mathbf{A}_{\left(n-i\right)\times k})\right)-k},\nonumber
\end{align}
where $\theta$ is chosen such that $\theta<\delta$.
For each
$i\in\left(n\left(\epsilon-\theta\right),n\left(\epsilon+\theta\right)\right)$, there is an $\alpha\in\left(-\theta,\theta\right)$ for which $i=n\left(\epsilon-\alpha\right)$. Therefore, $n-i=n\left(1-\epsilon+\alpha\right)>n\left(1-\epsilon-\delta\right)=k$. Now, according to Lemma \ref{lem:ref}, if we substitute $k$ for $\mathbb{E}\left(rank(\mathbf{A}_{(n-i)\times k})\right)$, as $n\to\infty$, we can write
\begin{align}
\mathbb{E}_{\mathbf{A}}(p_c(\mathbf{A}))\ge&\sum_{i=n(\epsilon-\theta)}^{n(\epsilon+\theta)}\binom{n}{i}{\epsilon}^i
{(1-\epsilon)}^{n-i}2^{\mathbb{E}\left(rank(\mathbf{A}_{\left(n-i\right)\times k})\right)-k}\nonumber\\
&=\sum_{i=n(\epsilon-\theta)}^{n(\epsilon+\theta)}\binom{n}{i}{\epsilon}^i
{(1-\epsilon)}^{n-i}2^{k\left({\frac{\mathbb{E}\left(rank(\mathbf{A}_{\left(n-i\right)\times k})\right)}{k}-1}\right)}\nonumber\\&\doteq
\sum_{i=n(\epsilon-\theta)}^{n(\epsilon+\theta)}\binom{n}{i}{\epsilon}^i{(1-\epsilon)}^{n-i}
\doteq\sum_{i=0}^{n}\binom{n}{i}{\epsilon}^i{(1-\epsilon)}^{n-i}=1.\nonumber
\end{align}
Therefore, $\lim_{n\to\infty}\mathbb{E}_{\mathbf{A}}(p_c(\mathbf{A}))=1$.
$\blacksquare$
\section{\textbf{The proof of Theorem \ref{Th:BECrow}}}
\textbf{\emph{Proof of Theorem \ref{Th:BECrow}}}:

According to the proof of Theorem \ref{BECgeneral} it is sufficient to show the following lemma.
\begin{lemma}
Suppose $\delta\ge 0$ and consider $\mathcal{R}_{n(1-\delta)\times n}$ with its previoiusly defined distribution. Then for $\mathbf{A}_{n(1-\delta) \times n}\in \mathcal{R}_{n(1-\delta)\times n}$ we have $\mathbb{E}\left(rank\left(\mathbf{A}_{n(1-\delta) \times n}\right)\right)\approx n(1-\delta)$. 
Note that here rank is calculated in $\mathbb{GF}(2)$.
\end{lemma}
\emph{Proof:} In order to prove the lemma it is sufficient to show that
\begin{equation}
\lim_{n\to \infty}\mathbb{P}\left(rank\left(\mathbf{A}_{n(1-\delta)\times n}\right)=n(1-\delta)\right)=1,\nonumber
\end{equation}
which is equivalent to show that the probability of having a matrix $\mathbf{A}_{n(1-\delta)\times n}$ with linear dependent rows goes to zero as n approaches inifinity, i.e.,  
\begin{equation}\label{Eq:tem1}
\lim_{n\to\infty}\left(\sum_{k=1}^{n(1-\delta)}\binom{n(1-\delta)}{k}\mathbb{P}(A_1+A_2+...+A_k=0)\right)=0,
\end{equation}
where $A_i$ represents the $i^{th}$ row of the matrix.
Suppose $\zeta$ be a positive number such that $\zeta<\rho$. The summation of equation (\ref{Eq:tem1}) can be written as
\begin{align}\label{Eq:tem2}
\sum_{k=1}^{n\zeta}\binom{n(1-\delta)}{k}\mathbb{P}(A_1+A_2+...+A_k=0)+\sum_{k=n\zeta}^{n(1-\delta)}\binom{n(1-\delta)}{k}\mathbb{P}(A_1+A_2+...+A_k=0).
\end{align}
We first prove that the first term tends to zero. In order to have $A_1+A_2+...+A_k=0$, $A_k$ should be equal to the sum of $A_1$ to $A_{k-1}$. Thus, conditioning on $A_1$ to $A_{k-1}$, it is easy to see that $\mathbb{P}(A_1+A_2+...+A_k=0)\le \frac{1}{\binom{n}{\rho n}}$. Thus,
\begin{align}
&\sum_{k=1}^{n\zeta}\binom{n(1-\delta)}{k}\mathbb{P}(A_1+A_2+...+A_k=0)\le\left(\sum_{k=1}^{n\zeta}\binom{n}{k}\right) \frac{1}{\binom{n}{\rho n}}.\nonumber
\end{align}
Since $\binom{n}{n\rho}\approx \frac{2^{h(\rho)n}}{(\rho(1-\rho)n2\pi)^{1/2}}$, we have
\begin{equation}
\lim_{n\to\infty}\left(\sum_{k=1}^{n\zeta}\binom{n}{k}\right) \frac{1}{\binom{n}{\rho n}}\le \lim_{n\to\infty}\frac{n\zeta 2^{n h(\zeta)}2^{-n h(\rho)}}{(\zeta(1-\zeta))^{1/2}(\rho(1-\rho))^{-1/2}}=0,\nonumber
\end{equation}
where in the last equality we used the fact that $h(\zeta)<h(\rho)$.

in order to complete the proof it is sufficient to show that the second term of equation (\ref{Eq:tem2}) 
also goes to zero. 
In this regard we show that for sufficiently large $n$ and any $k\ge\zeta n$ we have
\begin{equation}\label{Eq:randomwalk}
\mathbb{P}(A_1+A_2+...+A_k=0)\le \left(\frac{1}{2^{1-\frac{\delta}{2}}}\right)^n.
\end{equation}
This will prove the lemma because we would have
\begin{align}
&\lim_{n\to\infty}\sum_{k=n\zeta}^{n(1-\delta)}\binom{n(1-\delta)}{k}\mathbb{P}(A_1+A_2+...+A_k=0)\le\nonumber\\&
\lim_{n\to\infty}\left(\sum_{k=n\zeta}^{n(1-\delta)}\binom{n(1-\delta)}{k}\right)\left(\frac{1}{2^{1-\frac{\delta}{2}}}\right)^n
\le \lim_{n\to\infty}2^{n(1-\delta)}\left(\frac{1}{2^{1-\frac{\delta}{2}}}\right)^n=0.\nonumber
\end{align}

In order to prove equation (\ref{Eq:randomwalk}) we employ coupling method from random walk theory. Consider a random walk on the $n$-dimensional cubic in $\mathbb{GF}(2)$ with the set of directions $S$, consists of all $n$-dimentional vectors with $\rho n$ ones. Suppose this random walk starts from the origin, and each time selects its next direction randomly from $S$ with uniform distribution. Therefore, $\mathbb{P}(A_1+A_2+...+A_k=0)$ represents the probability of returning back to the origin after $k$ steps. Denote this random walk by the sequence $\{X_t\}$ of $n$-dimentional vectors where $X_t$ represents the position of the random walk after $t$ steps. Note that the stationary distribution of this random walk is uniform distribition, which means as $t$ tends to infinity the probability of being at any points of the cubic is almost $(\frac{1}{2})^n$. Thus, for large values of $k$, $\mathbb{P}(A_1+A_2+...+A_k=0)$ is almost $(\frac{1}{2})^n$. Now Consider another random walk denoted by $\{Y_t\}$, which its starting point is selected randomly with the uniform distribution. The idea of coupling is to couple two random walks $\{X_t\}$ and $\{Y_t\}$ with the dependency between the directions selected by them such that both of them remain random walks that select their directions in each step uniformly form  $S$. Suppose $X_i$ and $Y_i$ are the positions of the two random walks after $i$ steps and $s_{i+1}^x$ be the $(i+1)^{th}$ direction which is selected uniformly from $S$ by the random walk $\{X_t\}$. Suppose $r_i$ entries of the vectors $X_i$ and $Y_i$ are the same and denote the positions of these entries by the set $U=\{u_1, u_2, ... , u_{r_i}\}$. Let $S_U$ be the subset of $S$ consists of vectors that their $r_i$ entries with positions from $U$ are same as $s_{i+1}^x$. The random walk $\{Y_t\}$ select the direction $s_{i+1}^y$ uniformly from the set $S_U$. Note that due to the fact that $\{Y_t\}$ starts with its stationary distribution the probability of being at any point remains uniform for all $t$ for this random walk. Also note that according to the dependency between $s_{i+1}^x$ and $s_{i+1}^y$, $\{r_i\}$ is a non-decreasing sequence. Thus, we expect that the two random walk meet each other at a point. Let $\tau$ be the first time that $\{X_t\}$ and $\{Y_t\}$ meet. Note that after $\tau$ the rest of the two random walks would be the same. Conditioning on the $\tau$, $\mathbb{P}{(X_k=0)}$ can be written as
\begin{align}
\mathbb{P}(X_k=0)=\mathbb{P}(X_k=0|\tau\le k)\mathbb{P}(\tau\le k)+\mathbb{P}(X_k=0|\tau>k)\mathbb{P}(\tau>k).\nonumber
\end{align}
Now if we can prove that for $k\ge\zeta n$, $\mathbb{P}(\tau> k)$ goes to zeros as n tends to infinity, then we would have
\begin{align}
&\lim_{n\to\infty}\mathbb{P}(X_k=0)=\lim_{n\to\infty}\mathbb{P}(X_k=0|\tau\le k)\mathbb{P}(\tau\le k)+\mathbb{P}(X_k=0|\tau>k)\mathbb{P}(\tau>k)\nonumber\\
&=\lim_{n\to\infty}\mathbb{P}(Y_k=0|\tau\le k)\mathbb{P}(\tau\le k)+\lim_{n\to\infty}\mathbb{P}(X_k=0|\tau>k)\mathbb{P}(\tau>k)\nonumber\\
&=\lim_{n\to\infty}(\frac{1}{2})^n\mathbb{P}(\tau\le k),\nonumber
\end{align}
which proves the equation (\ref{Eq:randomwalk}). Note that $\mathbb{P}(\tau\le k)$ tends to 1 as k approaches infinity. Therefore to complete the proof it remains to show that 
\begin{equation}
\lim_{n\to\infty}\mathbb{P}(\tau> \zeta n)=0.
\end{equation}

For $1<j\le n$, suppose ${\tau}_j$ represents the first time that the $j^{th}$ entries of $X_t$ and $Y_t$ become the same. Thus we have
\begin{equation}
\mathbb{P}(\tau> \zeta n)\le \sum_{j=1}^n \mathbb{P}({\tau}_j> \zeta n)=n\mathbb{P}({\tau}_1> \zeta n)  .\nonumber
\end{equation}

Suppose that after $i$ steps the first entries of $X_i$ and $Y_i$ are not the same and let $r_i$ and the set $U$ be as defined previously. Let $\rho<\frac{1}{2}$. The first entry of $s_{i+1}^x$ is equal to one with probability $\rho$. Now due to the fact that $\rho<\frac{1}{2}$, less that $\frac{n-r_i}{2}$ entries of $s_{i+1}^x$ which are not from $U$ are equal to one with a probability more than $\frac{1}{2}$. This means that the first enrty of $s_{i+1}^y$ is equal to zero with a probability more than $\frac{1}{4}$. Thus, the first entries of $s_{i+1}^x$ and $s_{i+1}^y$ are not the same with a probability more than $\frac{\rho}{4}$. A similar approach for the case $\rho \ge \frac{1}{2}$ shows that there is a positive probability $p$ independent from $n$ and $i$ such that the first entries of $s_{i+1}^x$ and $s_{i+1}^y$ differ, i.e., the first entries of $X_{i+1}$ and $Y_{i+1}$ are the same. Thus we have 
\begin{equation}
\lim_{n\to\infty}\mathbb{P}(\tau> \zeta n)\le \lim_{n\to\infty}n\mathbb{P}({\tau}_1> \zeta n)\le \lim_{n\to\infty}n(1-p)^{n\zeta}=0.\nonumber
\end{equation}

This completes the proof. $\blacksquare$
\newpage
\section*{Acknowledgment}
The authors would like to thank Professor G. D. Forney for his valuable comments and suggestions and Mr. R. Farhoudi for his comments about proof of theorems.
\bibliographystyle{IEEEtran}


\end{document}